\documentclass[aps,prb,twocolumn,showpacs]{revtex4}

\usepackage{graphicx}
\DeclareGraphicsExtensions{.png,.jpg,.eps}

\usepackage{xcolor}
\usepackage{color}
\usepackage{amsmath}
\usepackage{url}

\newcommand{\redmark}[1] {#1}

\begin{document}

\title{Edge states in  graphene-like systems }

\author{J. L. Lado$^{1,}$\footnote{Tel. number: +351 253 140 112,\newline 
email address: jose.luis.lado@gmail.com}
, N. Garc\'ia-Mart\'inez$^1$, J. Fern\'andez-Rossier$^{1,2}$
}
\affiliation{$^1$International Iberian Nanotechnology Laboratory (INL),
Av. Mestre Jos\'e Veiga, 4715-330 Braga, Portugal
}
\affiliation{$^2$Departamento de Fisica Aplicada, Universidad de Alicante, 03690 Alicante, Spain}

\author{}

\date{\today} 

\begin{abstract} 

The edges of graphene and graphene like systems  can host localized   states  with evanescent wave function with properties radically different
from those of the  Dirac electrons in bulk.  
 This happens in a variety of situations, that are  reviewed here.
First,  zigzag edges host a set of localized non dispersive state at the Dirac energy. At half filling,  it is expected that these states are prone to ferromagnetic instability, causing a very interesting type of edge ferromagnetism.    Second, graphene under the influence of external perturbations can host a variety of topological insulating phases, including the conventional Quantum Hall effect, the Quantum Anomalous Hall (QAH) and the Quantum Spin Hall phase, in all of which  phases conduction can only take place through topologically protected edge states. 
 Here we provide an unified vision of  the properties of all these  edge states, examined under the light of the same  one orbital tight-binding model. We consider the combined action of interactions, spin orbit coupling and magnetic field, which produces a wealth of different physical phenomena. We  briefly address what has been actually observed experimentally. 
\begin{center}
\end{center}
  
\end{abstract}
\maketitle

\section{Introduction}
Graphene has been the most studied material of the last decade.  Its extraordinary electronic and mechanical properties came as a great surprise: the existence of stable two dimensional crystals was customarily dismissed,  and surfaces had been identified as the source of reduction of electronic mobility, due to  defects and adsorbate trapping. The  age of graphene was initiated by the observation of the \redmark{field} effect transistors\cite{Novoselov04}, and more strikingly \redmark{the} quantum Hall effect\cite{Geim05,Kim05},  a  phenomena that had only been observed in high mobility semiconductor heterostructures\cite{VK}.  

Graphene is a two dimensional lattice of carbon atoms that  form a honeycomb lattice, that can also be described as  a triangular lattice with a two atom basis,    displayed with different colors in \redmark{Fig} 1.  This makes of the graphene honeycomb lattice  a bipartite lattice, a fact that strongly influences its electronic properties.
The electronic properties of graphene  can be described in terms
of a very elegant and simple picture\cite{RMP07,Katsnelsonbook}
\redmark{by means of the Dirac equation}.
Close to the Fermi energy, electrons in graphene behave as two dimensional relativistic masless particles, the so called Dirac electrons. The energy bands are linear,  $E_{\pm}=\pm \hbar v_F |\vec{k}|$, so that the  three dimensional plot of these  two dimensional bands produces the so called Dirac cones.
  The Brillouin zone associated to the honeycomb lattice is also hexagonal, and has a copy of  these Dirac bands, located at the corners of the hexagon. Only two of these so called valleys  are actually non-equivalent.
 As a result,  electrons in graphene have an additional isospin,  the valley. 
 
  All these properties are also expected for a wider class of material systems,  the graphene-like materials,  that can also be described in terms of electrons moving in a honeycomb lattice with just one orbital per site.  An incomplete list of graphene-like materials includes  Silicene\cite{glike1}, Germanene\cite{glike2}, Stanene\cite{glike3}, Metallic organic framework\cite{glike4},  hydrogenated Bi(111)\cite{Niu15}, and artificial graphene lattices\cite{glike5}. 
 
The purpose of this paper is to review what is known about  the fate of the Dirac electrons at the edges, the boundaries of these otherwise endless two dimensional  crystals.  In some instances Dirac  electrons  simply scatter at the edges but, very often,  graphene hosts edge states, {\em i.e.}, states  whose wave function are evanescent in the direction perpendicular to the edge, and itinerant in the  parallel direction.  Their energies are at, or close to, the  Dirac point,  and very often their wave functions have peculiar properties, such as sublattice polarization,  spin polarization or net spin current, just to mention a few. 

Edge states are particularly important when graphene is driven into what nowadays are known as topological insulator phases\cite{Shenbook}.  Historically,  the first example of this phase is associated to the Quantum Hall Effect (QHE)\cite{VK}, observed in high mobility two dimensional electron gases in semiconductor heterostructures.  In these systems, application of a sufficiently large magnetic field produces a discrete spectrum of Landau levels  (LL) in the bulk states. This leads to an  insulating state when the Fermi energy lies in between the LL.  Importantly, in that situation the edges host   chiral (or unidirectional) states that are ideal quantum conductors \cite{Laughlin81,Halperin82}, and are responsible of the perfect quantization of the Hall conductance\cite{VK}.  

Soon after the discovery of the QHE,  it was shown  by Thouless and coworkers (TKNN\cite{TKNN}) that the Hall conductance could be expressed, using the conventional linear response theory,  in terms of a topological invariant\cite{TKNN,Thouless-book}, the so called Chern number ${\cal C}$,  associated to the Berry curvature of wave functions of the bulk states.  

The prediction of other insulating phases with quantized edge transport due to  topological order without a net  magnetic field is one of the greatest successes of  modern condensed matter theory.  This includes  the quantum spin Hall (QSH)\cite{Kane-Mele1,Kane-Mele2}, and the quantum Anomalous Hall (QAH) phases, proposed  in a  seminal paper by Haldane\cite{Haldane88}  where he showed how spinless fermions  moving in a honeycomb lattice exposed to a periodic magnetic field with no net flux   would display quantized Hall conductance, with topologically protected edge states.  
 \begin{figure}
 \centering
  \includegraphics[width=0.5\textwidth]{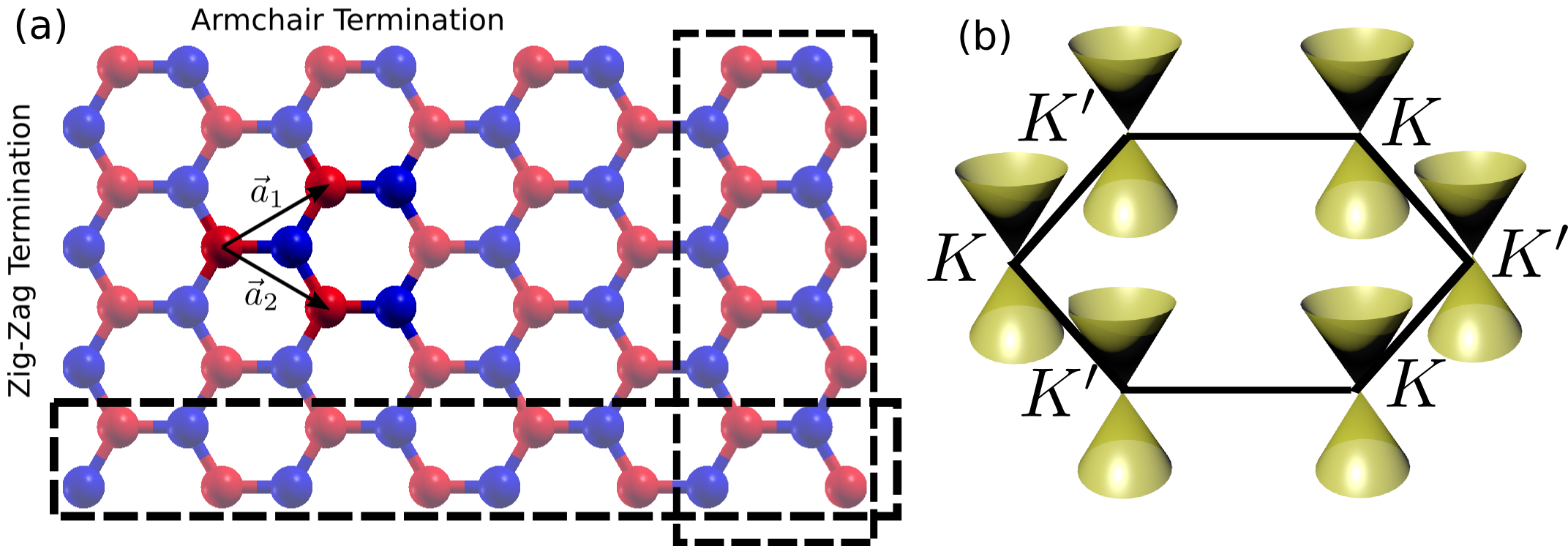}
\caption{(a) Honeycomb lattice showing two types of edge, zigzag and armchair.  The two triangular sublattices, $A$ and $B$, are displayed with fake color, red and blue.  The vectors of the Bravais lattice are also shown. (b) Brillouin zone associated to the honeycomb lattice, including the plot of the two  energy bands forming Dirac cones  in the neighborhood of $K$ and $K'$ points (see text)  }
\label{f1}
\end{figure}

The QSH phase was proposed  by Kane and Mele\cite{Kane-Mele1,Kane-Mele2}.  They found that intrinsic spin-orbit coupling would open a gap  in  graphene with non-trivial topological order that would come accompanied by spin filtered\cite{Kane-Mele1}  edge states robust with respect to time reversal perturbations.    Interestingly, the description of electrons with  spin-orbit coupling in graphene was mathematically identical to two independent copies of the Haldane model, one per spin.  
They also introduced a $Z_2$ topological classification\cite{Kane-Mele2}
of time-reversal  invariant two dimensional systems, analogous to the TKNN
classification of quantum Hall states. 

Subsequent computational work\cite{Min06,Yao07} showed that the magnitude of the intrinsic spin-orbit coupling  in graphene was so small   that would render the \redmark{observation of}
the QSH phase almost impossible. However, there are graphene-like materials, such as Silicene and other group IV honeycomb crystals,  for which the Kane Mele model applies\cite{Liu11} and for which these predictions are relevant.  More importantly, there is quite strong experimental evidence that the  QSH phase has been observed both in HgTe quantum wells\cite{Konig07} and inverted InAs/GaSb quantum wells \cite{Knez11}, both theoretically predicted to be QSH insulators\cite{Bernevig06,Liu08}


 The role of  Coulomb interactions can also affect dramatically the properties of some of these edge states, in particular, whenever  edge states produce
 a large density of states at the Fermi energy, that makes them prone to Stoner instabilities. This is the case of zigzag edge states for which  ferromagnetic order is expected\cite{Fujita96,Waka98,Son06,Cohen06,JFR07,Gunlycke07,JFR08,Kim08,Munoz09,Yazyev10}.  The interplay between this magnetism, spin-orbit interactions\cite{Sandler07,Soriano10,Gosalbes2011,Lado14a} and the Quantum Hall phases\cite{Lado14b} is a very fascinating area of research that we also review here. 

\redmark{
Appart from the previous examples, interfacial effects can also create
topologically protected states. Some examples are driven by domain
boundaries between gapped graphene\cite{jackiw,Lado13}, local electric
edge fields\cite{armchair-state1,armchair-state2} or interfaces between
antiferromagnetic graphene and a superconductor.\cite{Pablo2015} }

The rest of this review is organized as follows. In section II we review the tight-binding model that describes both the 2D and edge states in graphene, including the spin-orbit coupling and coupling to the magnetic field, responsible of the Quantum Spin Hall and Quantum Hall phases.   In section III we review the properties of the zigzag edge states, including their connection with the bipartite character of the honeycomb lattice as well as the ferromagnetic order associated to Coulomb interactions.  In section IV,V and VI we review the quantum Hall, quantum anomalous Hall and quantum spin Hall  edge states, respectively.  In sections VII and VIII we review the effect of Coulomb interactions on the spin-filtered edge states in graphene.  In section IX we briefly review the experimental situation and in section X we wrap up  with some general conclusions.

\section{ Tight binding model for graphene and graphene-like materials}
A material is said to be graphene-like if  its quantum states can be described in terms of  a tight-binding model that describes electrons hopping in a honeycomb lattice with a single state per site. In the case of graphene, Silicene, etc,  the site would be a group $IV$ atom,  and the state  would be  $p_z$ orbital.    Within this model, electrons can hop to their first neighbor atoms, with a hopping amplitude $t$, that takes a value of $t\simeq 2.7$eV\cite{RMP07} in the case of graphene.    
The tight-binding approach  can also include the  effect of magnetic fields
by means of the so called Peierls substitution. Basically, the effect of the magnetic field consists on
multiplying by a
phase the hopping integrals $t_{\alpha,\beta}\rightarrow t_{\alpha,\beta}e^{i\phi_{\alpha,\beta}}$ where
\begin{equation}
 \phi_{\alpha,\beta} = \frac{e}{\hbar}\int^{\beta}_{\alpha}\vec{A} \cdot d\vec{r}
 \label{Peierls}
\end{equation}
and  $\vec{A}$ is the vector potential applied to the system.

\subsection{Bloch and Dirac Hamiltonians}
The honeycomb lattice of \redmark{bulk}
graphene can be treated as  two interpenetrating  triangular lattices,  that we label as $A$ and $B$ and assign them the red and blue color
in \redmark{Fig. 1}. 
Thus, the honeycomb lattice is a triangular lattice  with two atoms per unit cell that naturally leads, within the simple TB model,  to a Bloch Hamiltonian of dimension two:
\begin{eqnarray}
{\cal H}_0(\vec{k})= \left(
\begin{array}{cc}
\frac{\Delta}{2} & t f(\vec{k}) \\
t f^*(\vec{k}) & -\frac{\Delta}{2}
\end{array}
\right)
\end{eqnarray}
where  $t$ is the first neighbor hopping, $\Delta$ is the so called mass term that is present whenever there is a sublattice symmetry breaking perturbation and it is assumed to  vanish in the case of freestanding  graphene, 
\redmark{$f(\vec{k})=1+e^{i\vec{k}\cdot\vec{a}_1}+e^{i\vec{k}\cdot\vec{a}_2}$}
is the form factor associated to first neighbor hopping in the honeycomb lattice, and $\vec{a}_1=\frac{a}{2}\left(\sqrt{3},1\right)$, $\vec{a}_2=\frac{a}{2}\left(\sqrt{3},-1\right)$
where $a$ is the unit cell spacing, which coincides with the second neighbor distance and it satisfies the relation $a=\sqrt{3}a_{CC}$ with  the first neighbor distance. 
The resulting energy bands, $\epsilon_{\pm}(\vec{k}) = \pm  \sqrt{|\frac{\Delta}{2}|^2+|t f(\vec{k} |^2}$ are shown in
Fig. \ref{f1}b for the $\Delta=0$ case relevant for graphene , and feature
the so called Dirac cones at the corners of the hexagonal Brillouin
Zone.  Valence and conduction band meet at the so called Dirac
point, which coincides with the Fermi energy at half filling. 
At this point we introduce the concept of sublattice as a pseudo spin
degree of freedom. For that matter, we can write down the Bloch
Hamiltonian in terms of the Pauli matrices $\sigma_x,\sigma_y,\sigma_z$: 
\begin{eqnarray}
{\cal H}_0(\vec{k})= 
t \left [ Re(f(\vec{k})) \sigma_x +
Im(f(\vec{k})) \sigma_y \right ]
 + \frac{\Delta}{2}
\sigma_z 
=\vec{h}\cdot\vec{\sigma}
\end{eqnarray}
This notation makes it apparent that the Bloch states of graphene,
that we can describe as two component spinors
$\left(\begin{array}{c} \phi_A\\ \phi_B\end{array}\right)$.
Actually, for
$\Delta=0$, those states
 are analogous to the eigenstates of a
pseudo spin under the influence of an in-plane magnetic field.

As a result,  the projection of their wave functions over the two sublattices have the same weight $|\phi_A|=|\phi_B|$. In other words,  the {\em bulk} states of graphene are sub-lattice unpolarized. 

In the neighborhood of the $K$ and $K'$ points,  denoted by the label $\tau_z=\pm 1$, it is convenient to Taylor expand the Bloch Hamiltonian to obtain the so called Dirac Hamiltonian:
\begin{eqnarray}
{\cal H}_{0}(\vec{q})= \hbar v_F \left(q_x \sigma_x  + \tau_z q_y \sigma_y\right) + \frac{\Delta}{2}\sigma_z
\label{kp}
\end{eqnarray}
where $\vec{q}\equiv \vec{k}- \vec{K}_{\tau}$ and $v_F=3 t a_{CC}/2 \hbar$,   Many of the low energy properties of graphene can be understood
in terms of the previous continuous model.


\subsection{Spin dependent terms}
The spin plays a crucial role in most of the edge state physics in graphene and graphene-like systems. 
The Hamiltonian of electrons in graphene can have up to three different  spin-dependent terms.  At finite field, the Zeeman term
\redmark{
\begin{equation}
{\cal H}_Z =\frac{1}{2}g\mu_B \vec B \cdot \vec \sigma
\label{Zeeman}
\end{equation}
}
where $g\simeq 2$. 

The other term is spin orbit
coupling (SOC). The original atomic spin-orbit
term, $\lambda \vec{S}\cdot\vec{L}$, has a vanishing value on the $p_z$.
However, higher order processes involving orbitals
from the $p_x$ and $p_y$ manifold, or even from the $d$
manifold,\cite{graphene-dorb}
will add an effective SOC term to the Hamiltonian.

Whereas a constructive procedure has not been
derived in general,  the following Hamiltonian postulated by Kane
and Mele,  reproduces all the important properties that the
actual SOC  brings in more realistic calculations\cite{Kane-Mele1,Kane-Mele2}: 
\begin{equation}
H_{KM} = 
\displaystyle\sum_{\langle\langle \alpha,\beta\rangle\rangle,\sigma}
it_{KM} \sigma\nu_{\alpha,\beta}c^{\dagger}_{\alpha\sigma}c_{\beta\sigma}
\label{Hamil}
\end{equation}
 double angle brackets denote second neighbors summation, $\sigma=\pm1$ are the spin projections (along the axis perpendicular to the crystal plane) and $\nu_{\alpha\beta}=+(-)1$ for clockwise (anticlockwise) second neighbor hopping. 
 
   When added to the first-neighbor hopping Hamiltonian, the  Kane-Mele term opens a band-gap $\Delta_{SOC}= 6\sqrt{3}t_{KM}$  at the Dirac points.   Thus, SOC would turn graphene into a gapped material and, as it was also found out by Kane and Mele, of a very special nature.  In contrast with
most  of the band insulators, graphene was predicted to be a quantum Spin
Hall insulator, i.e., topologically different from
vacuum.  A dramatic consequence of the topological non-triviality of the QSH phase is the existence of chiral edge states in graphene.

At first sight, it is somewhat surprising that 
in the Kane-Mele  spin-orbit coupling Hamiltonian,
$S_z$ is a good quantum number,  
$\left [ {\cal H}_{KM} , S_z \right ] = 0$.
Mirror symmetry is the ultimate cause of this conservation law, which can also be understood as follows. Spin-flip terms $S^{+} L^{-}+ S^{-}L^{+}$ connect the $\pi$ orbitals with the $p_{x,y}$ orbitals. Thus, the effective Hamiltonian has to include these
processes in couples, so that  the $\pi$ electron with spin $\uparrow$ couples to a state $p_x+ip_y$ with spin $\downarrow$ and then
return to the original state, preserving the spin thereby. In addition,
density function theory calculations shows\cite{graphene-dorb} that
even though the low energy
properties are dominated by the $p_z$ orbitals, the SOC contribution
comes also from the $d$ manifold. Actually, this
contribution turns out to be
even bigger than the one from the $p$ manifold.

Whenever  mirror symmetry is broken,  due to application of an external off-plane electric field, or due to interaction with the substrate,  another spin orbit coupling arises known as the Rashba Hamiltonian\cite{Kane-Mele1,Min06}:
\begin{equation}
{\cal H}_R = it_R \sum_{i,j,s,s'} \vec E \cdot \left ( \vec r_{ij} \times \vec \sigma \right )_{s,s'}
c_{is}^\dagger c_{js'}
\label{Rashba}
\end{equation}
where $\vec r_{ij}$ is unit vector along the bond  between the carbon sites $i$ and $j$, 
$\vec \sigma$ are the spin Pauli matrices and $\vec{E}$ is a vector related to inversion symmetry breaking of the graphene lattice, such as an off-plane electric field\cite{Min06}. The Rashba spin orbit coupling does not commute with $S_z$. 

\subsection{Coulomb interaction}
In general, electron-electron interactions play a secondary role in the electronic properties of graphene and graphene-like systems\cite{RMP-int}, whose defining properties are captured by the tight-binding model presented above.
However, in those instances where the single-particle spectrum has a large degeneracy, such as the case of zigzag edge states as well as the bulk LL, interactions can have a strong effect.  
In order to model electron-electron interactions,  two approximations are often employed. First, only the intra-atomic part of the interaction is considered. This is the so called Hubbard approximation: 
\begin{equation}
{\cal H}_{U} = U\displaystyle\sum_{i}n_{i\uparrow}n_{i\downarrow}
\end{equation}
where the sum runs over the sites $i$ of the lattice. 
The resulting Hubbard model can not be solved  exactly, except in  a monostrand one dimensional chain. Thus, very often the model is treated at the mean field approximation, where the exact Hamiltonian is replaced by an effective Hamiltonian
\begin{equation}
{\cal H}_{MF} = {\cal H}_{\rm Hartree} + {\cal H}_{\rm Fock}
\end{equation}
where 
\begin{equation}
{\cal H}_{\rm Hartree} = 
U \left( n_{i,\uparrow}\langle n_{i,\downarrow} \rangle
+
n_{i,\downarrow} \langle n_{i,\uparrow} \rangle
\right)
\label{Hartree}
\end{equation}

\begin{equation}
{\cal H}_{\rm Fock } = 
-U\left(
c_{i,\downarrow}^\dagger c_{i,\uparrow}
\langle
c_{i,\uparrow}^\dagger c_{i,\downarrow}
\rangle
+
c_{i,\uparrow}^\dagger c_{i,\downarrow}
\langle
c_{i,\downarrow}^\dagger c_{i,\uparrow}
\rangle
\right)
\label{Fock}
\end{equation}
so that electrons interact with an external field that is
self-consistently calculated.  In most of papers\cite{Fujita96,Waka98,JFR07,JFR08,JeilJung09,Munoz09,Yazyev10,Soriano10,Yazyev11b} an additional approximation has been used, that assumes  a collinear magnetization so that the Fock term vanishes. However, instances in which this is not the case are very interesting and also reviewed here\cite{Lado14b}.

\subsection{Calculation method for edge states}
Two methods are normally  used to compute edge states starting from a tight-binding Hamiltonian.   In most instances  we compute the energy bands of a so called graphene  ribbon,  a  one dimensional crystal  with two edges  \cite{Nakada96}.  When the width of the ribbon is sufficiently large, compared to the penetration length of the edge states,  the interactions between the edges are  negligible, allowing to study the properties of the edge states.  Graphene ribbons are interesting by their own sake, and there has been enormous progress in the fabrication of ribbons with smooth edges\cite{Dai08,Dai10,Dai11,Crommie2011,Magda14}, so that inter-edge coupling is also an interesting topic\cite{Brey-Fertig, Waka98,Son06,Cohen06,JFR08,JeilJung09,Munoz09,Yazyev10,Soriano10,Lado14a,Magda14}. 

 A second strategy to calculate edge states is to calculate the Green's function of a semi-infinite two dimensional crystal using the recursion method.  Using the translation invariance along the direction parallel to the edge, it is possible to write the Hamiltonian of the semi-infinite crystal as a one dimensional semi-infinite crystal whose effective Hamiltonian depends on the transverse wave vector $k$.   By so doing, the  conventional Green's function techniques\cite{Dattabook} normally used to deal with 1D problems can be used. In particular, the Green's function of the unit cell in the semi-infinite crystal reads:
 \begin{equation}
G (k,E) = \left ( E - h_0(k) - \Sigma(k,E) + i\epsilon      \right )
\label{green1}
\end{equation}
where $\Sigma(k)$ is the self-energy induced by the coupling to the rest of the crystal: 
\begin{equation}
\Sigma (k,E) = t(k)G (k,E) t^\dagger(k)
\label{self1}
\end{equation}
where $h_0(k)$ and $t(k)$ are the intracell and intercell matrices of the Block Hamiltonian. Equations (\ref{green1}) and (\ref{self1}) define a set of non-linear coupled matrix equations that is solved by numerical iteration.   Once the surface Green's function  is derived, the density of states can be obtained through $\rho(k,E)= -\frac{1}{\pi}{\rm Im}\left(G(k,E)\right)$.  Contour plots in the $k,E$ plane can reveal the existence of in-gap edge states, as shown below. 


Most of the results presented in this manuscript reproduce existing results of the literature. However, we have re-computed most of them using a home-made code, \textsc{Quantum Honeycomp},  a package that  computes the electronic properties of honeycomb tight-binding models in one dimension, including the couplings to the magnetic field (eqs. (\ref{Peierls},\ref{Zeeman}), the   spin-orbit couplings (eqs. (\ref{Hamil},\ref{Rashba})) as well as the mean field Hubbard terms in the non-collinear approximation (eqs. (\ref{Hartree},\ref{Fock}) ). The code is available online\cite{honeycomb} and has a graphical user interface that permits a simple use by non-experts. 

 \section{Zigzag edge states}  
 We now review the properties of the simplest type of edge state in graphene, the so called zigzag edge states. Figure 1a shows the two  simplest types of edges in graphene, the so called armchair and zigzag.  Of course, other terminations are possible in principle\cite{Yazyev11b}, but we shall not discuss them here.  As we show now, the zigzag terminations host $E=0$ energy states, whereas the armchair terminations do not.   
The ultimate reason for these states is sublattice polarization.
Firstly, the zigzag edges present a very strong sublattice imbalance.
Secondly
all the atoms of a given zigzag edge belong to the same sublattice. This turns out to be a very important property that leads to the existence of $E=0$ evanescent states. This can be seen in several ways. 

\subsection{Effective mass  description of Zigzag edge states}
A quick and simple argument for the existence of mid-gap $E=0$ edge states in graphene can be obtained using the effective mass Hamiltonian from eq. (\ref{Dirac-Landau-1}) taking $E=0$ and $B=0$ and proposing the antsaz $e^{i q y} \left(\begin{array}{c} \phi_0(x) \\ 0 \end{array}\right)$ in which the wave function lives only in one sublattice. It is apparent that $\phi_0(x)= e^{-q x}$ satisfies the Dirac equation.  Therefore, the localization length can be written as $\lambda =q^{-1}$, is maximal for the Dirac point, $q=0$, and decreases (the state becomes more localized) as $q$ increases. A more complete analysis along this line, including the calculation of the inter edge coupling in the case of graphene ribbons, can be found in the seminal work  of Brey and Fertig\cite{Brey-Fertig}.   The validity of the effective mass approximation to describe states  abrupt perturbations, such as an edge, is questionable a priori. However the results of this approximation compare well with the tight-binding results\cite{Brey-Fertig}.

\subsection{Tight-binding description of Zigzag edge states}
\begin{figure}[hbt]
 \centering
  \includegraphics[width=0.5\textwidth]{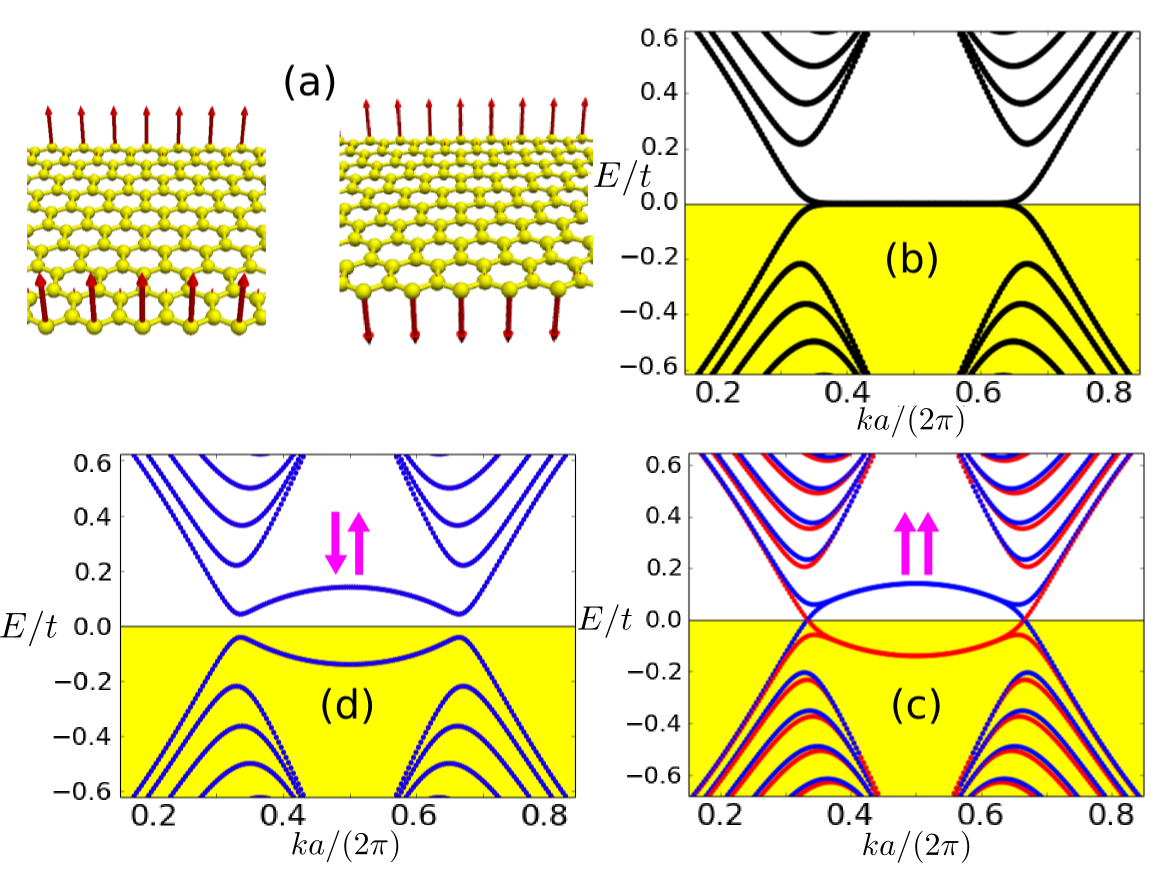}
\caption{(a) Calculated magnetic moments Zigzag edge states for a graphene ribbon with $N=20$ atoms in the unit cell, both in the AF and FM configurations, using the meant field Hubbard model with $U=t$.  Energy bands for the non-interacting (b),  and the AF (c) and FM (d) solutions for a ribbon with $N=40$ atoms in the unit cell. 
The color red (blue) denotes spin up (down), notice that in figure (a) only one of the spins are visible but the degeneracy is two for the whole spectrum.   }
\label{f2}
\end{figure}

The existence of an $E=0$ edge state was first inferred from the calculation of the energy bands of a zigzag graphene ribbon in the early work of Nakada and coworkers\cite{Nakada96}.  Direct diagonalization of the TB Hamiltonian yields the energy bands that we show in figure (\ref{f2}b).  Two types of bands are seen: confined bulk modes, that have a gap, and two flat bands with  $E\simeq0$ 
corresponding to states localized at the two  edges of the ribbon.
Small departures of their energy from $E=0$ are related to inter edge hybridization.    This occurs because the edge wave function associated to a given edge lives in the opposite sublattice than the edge state coming from the opposite edge.  An important property of the electronic states of zigzag ribbons is that
they do not mix valleys\cite{Brey-Fertig}. Actually, the confined bulk modes are cuts of the two Dirac cones, whereas the $E\simeq0$ flat bands  join the otherwise disconnected valleys. 

The $E=0$ wave function of a single edge can be worked out analytically\cite{Nakada96}.  If the edge atoms belong to the $A$ sublattice, they found that $\phi_B=0$ and $\phi_A(m)= \left(2 cos \left(\frac{ka}{2}\right)\right)^{2m}$ where $m=1,2,..$ labels the $A$ atoms starting from the edge along an armchair row.  This wave function is only normalized if $ka>\frac{2\pi}{3}$, i.e., 
between the Dirac points and the $\Gamma$ point,  which naturally explains the results of figure \ref{f2}b. An expansion of the wave function coefficient around the Dirac point   gives an exponential decay, similar to the one obtained using the Dirac Hamiltonian.


\subsection{Zero modes in bipartite lattices}
The zero modes of the zigzag edge can be  understood in a broader context. A one orbital tight-binding Hamiltonian defined on  any  bipartite lattice, 
with a 
number of atoms in one lattice larger than in the other ($N_A>N_B$), has  least
\begin{equation}
N_Z=|N_A-N_B|
\label{NZ}
\end{equation} 
states with zero energy whose wave function is localized in the $A$ (majority) sublattice\cite{Inui94}.  Three standard examples of systems with sublattice imbalance would be graphene with a vacancy\cite{Palacios08}, semi-infinite graphene with a zigzag edge and a triangular graphene island with zigzag edges\cite{JFR07}. 

 The proof of this theorem goes as follows.  For a bipartite lattice, the Schrodinger equation can be written as two linear systems: 
\redmark{
\begin{eqnarray}
\sum_{b} H_{ab}\Phi_b = E \Phi_a \nonumber \\
\sum_{a} H_{ba}\Phi_a = E \Phi_b 
\end{eqnarray}
}
In order to verify the theorem, we take $E=0$ and $\phi_b=0$ for the $N_B$ sites of the lattice, so that we are left with an undetermined linear homogeneous system of equations,  $\sum_{a} H_{ba}\Phi_a =0 $,  with $N_A$
variables, the components of the wave
function on the majority sublattice, but  only $N_B$ equations.  
There are an infinity of such solutions, which form a vector space, whose dimension is the difference between the number of unknowns $N_A$  and the rank of the matrix of the system, $N_B$.  This vector space is the one of the $E=0$ modes enunciated in the theorem.

In the case of a bipartite lattice with a constant staggered potential, such that all $A$ ($B$) atoms have an on-site energy $\frac{\Delta}{2}$ ($\frac{-\Delta}{2}$) the theorem can be also applied, but now the "zero modes' have energy $\pm\frac{\Delta}{2}$, depending on which one is the majority sublattice\cite{Soriano12}, and they are still entirely localized on a single sublattice.

\begin{figure}[hbt]
 \centering
  \includegraphics[width=0.5\textwidth]{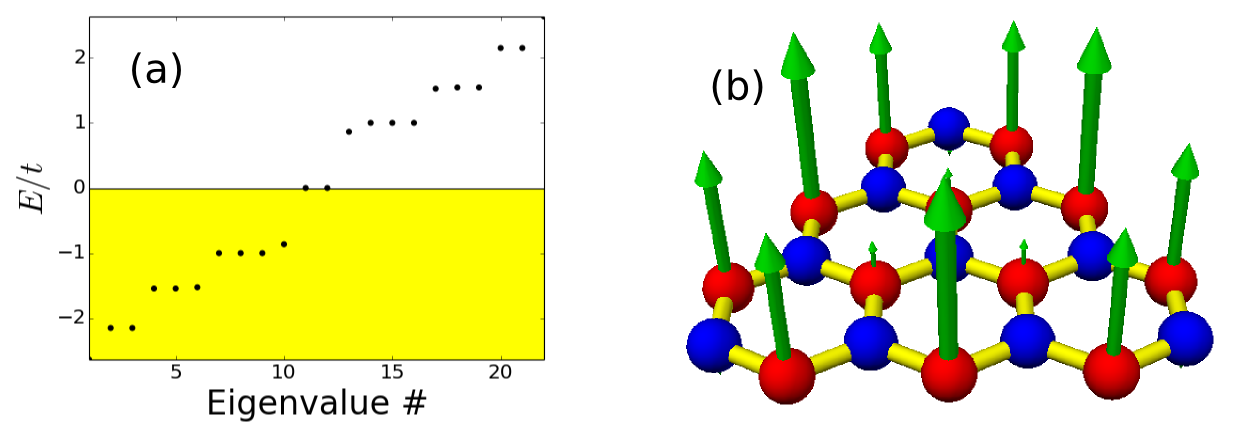}
\caption{(a) Non-interacting $(U=0)$ energy levels for the $N=22$ triangulene, Notice the two states with $E=0$. (b) Calculated magnetization, within the mean field Hubbard model with $U=t$.  The self-consistent solution has two unpaired electrons ($S_z=2$) whose magnetization goes predominantly to the majority sublattice atoms (shown in red) at the edges.}
\label{ftriang}
\end{figure}

A dramatic example of $E=0$ edge states is provided by triangular graphene nano islands with zigzag edges\cite{JFR07}, shown in figure \ref{ftriang}.  A remarkable feature of this class of systems is that their sublattice imbalance scales with size, and so it does the number of $E=0$ states. 
\redmark{So,} 
the triangular islands can have strict $E=0$  states, even for small systems. The smallest of such systems has an odd number of  $N=13$ atoms,  so that it has to have a sublattice imbalance, $N_A-N_B=1$.  But the second of such structures has an even number of atoms, $N=22$, which would permit to build many structures with $N_A=N_B$ such as pentancene , yet the triangulate has $N_A-N_B=2$ and, thereby two mid-gap $E=0$ states.  Inspection of the wave function reveal that their wave functions have more weight on the edges\cite{JFR07}.  Another example of $E=0$ edge state is provided by the minimal edge that can be devised, the one created by a single vacancy.  The theorem warrants the existence of a $E=0$ state. The lack of a gap in graphene turns  this $E=0$ state into a resonance\cite{} whose amplitude decays . In the case the quantum spin Hall phase, where a non-trivial gap is open,  the vacancy creates a real mid-gap state with a normalizable wave function\cite{Gonzalez12}.  In both cases, the vacancy localizes an electron

It must be stressed that the theorem of eq. (\ref{NZ}) gives the minimal number of zero modes of a given structure. As  discussed by Koshino and coworkers\cite{Koshino14}, the presence of symmetries in the Hamiltonian  that commute with the chiral symmetry, the one responsible of the electron-hole symmetry and other features of bipartite Hamiltonians, can increase the number of zero modes if the invariant subspaces associated to that symmetry have their own sublattice imbalance.  As a result, structures with a global null sublattice imbalance still have zero energy states\cite{Potasz10}.

\subsection{Edge magnetism: mean field Hubbard model}
The $E=0$ states discussed in the previous subsection play a very important role in graphene because, at half filling, the Fermi energy lies exactly at $E=0$.  Actually, in  a system with $N_A+N_B$ atoms at half filling, with $N_A>N_B$,  the number of bound states with $E<0$ is $N_B$, so that the valence band hosts $2 N_B$.  If we write the number of electrons $N=N_A+N_B= N_Z + 2N_B$, it is apparent that the in-gap $E=0$ states are half-full.   Given that the wave functions of the $E=0$ states overlap in space,  the Coulomb interactions are expected to favor ferromagnetic spin correlations, very much like in open shell atoms.   In a solid state physics parlance, the presence of edge states creates a large density of states at the Fermi energy. Therefore, Coulomb interactions could result in a Stoner instability that produces ferromagnetic order.

These hand-waving arguments can be put on a very firm basis.
Density functional theory (DFT) calculations
have shown that zigzag edges are indeed ferromagnetic\cite{Son06,Cohen06,JFR07}. 
In the case of graphene ribbons, the magnetization of the edges affects the otherwise flat edge bands,  which become dispersive.
When the magnetization of the edges is (anti)parallel, the
so called (anti)ferromagnetic configuration, the
ribbon is (non) conducting.  The change of the conduction
properties of a zigzag graphene ribbon with the relative
orientation of the edge magnetization inspired the proposal
of a graphene based spin valve\cite{Kim08,Munoz09} as well
as a Silicene spin valve\cite{Xu12}. 

Interestingly,  the description of the magnetic order in graphene ribbon using the mean field approximation of the Hubbard model\cite{JFR08,JeilJung09} gave results very similar to those of DFT and, in addition, provided analytic
 insight on origin of the magnetization induced dispersion as well as a treatable  description of the evolution of magnetism away from half filling\cite{Jung09b} .  The difference between the $U=0$ bands (fig. (\ref{f2}b)) and the $U\neq 0$ bands (fig. (\ref{f2}c,d)) is known as the  interaction self-energy.  In the case of the AF configuration, the self-energy is non zero only for the two edge bands, so that it can be calculated analytically\cite{JFR08}.  In that case the self-energy is a dimension two matrix and it has both intra-edge and inter-edge contributions. The former gives rise to the spin splitting of the band states close to the $\Gamma$ point $k=\pi$ and is thereby identical for the AF and FM solutions. 
As we move away from the $\Gamma$ point towards the valleys, the wave
functions become more delocalized and the inter-edge contribution
to the self-energy takes over and, depending on the 
relative orientation of the edge magnetic moments a gap
opens (AF) or a spin polarized  conducting channel remains
open  at each valley   (FM).

The results of the
mean field Hubbard model calculations for the  ferromagnetic (FM) and antiferromagnetic (FM) for a graphene ribbon are shown in figure (\ref{f2}), taking $U=t$ for a ribbon with $N=40$ atoms.  For this value of $U/t=1$ the
magnetization of the edge atoms is $0.15 \mu_B$,  three times smaller than the magnetic moment of Nickel atoms in ferromagnetic Nickel.   The magnetization of non-edge atoms is much smaller and decays exponentially as a function to the distance to the edge.  Within the same mean field Hubbard model, it would take a value of $U$ larger than $2.2t$, to produce magnetic order in the
\redmark{bulk} honeycomb lattice,  which would be antiferromagnetic (AF). More sophisticated approximations, such as quantum Monte Carlo,  push the critical value of $U$ to even
 higher values\cite{Sorella92}.   In the case of edge atoms, the mean field magnetization is non-zero for any positive value of $U$\cite{Fujita96}. 

In the case of triangular graphene islands, both density functional calculations and mean field Hubbard model calculations yield very similar results\cite{JFR07}. Triangular islands  whose single particle spectrum has $N_Z$ zero energy states develop edge magnetism with $S_Z=\frac{N_Z}{2}$, as if these systems were following the atomic Hund rule\cite{JFR07}.   Actually, in the case of the Hubbard model,  Lieb demonstrated the following theorem: the spin $S$ of the ground state of a Hubbard model defined in a bipartite lattice, at half filling,  is given by $S=\frac{N_A-N_B}{2}$.  Thus, the mean field approximation is, in some  sense, compatible with the Lieb theorem, but it is important to keep in mind the differences between the broken symmetry mean field solutions and the exact solutions.  In the case of graphene ribbons,  the AF configuration has lower energy than the FM solution, which very often is said to be in agreement with the Lieb theorem, but the exact solution with $S=0$ would have a vanishing magnetic moment at each edge, whereas the mean field AF solution the net magnetization vanishes, but not the local magnetization. 

\subsection{Beyond the mean field Hubbard model}
The mean field theory of the Hubbard model provides a handy first description of the magnetic properties of the edges of graphene and graphene-like materials.  However,  this approach has several well known shortcomings.  For instance, the long-range tail of Coulomb interactions in graphene should be poorly screened.  Therefore, non-local exchange effects are missed both by  mean field Hubbard approximation and DFT calculations based on local density approximations.  A proper treatment of non-local exchange\cite{Jeil11} shows that  magnetic features are further stabilized, both for intra-edge and inter edge exchange.

Both local and non-local mean field approximations ignore spin dynamics and the fact that long-range order is not possible in one dimension.   The spin waves of the magnetically ordered zigzag ribbon were calculated within the RPA approximation for the Hubbard model by Wakabayashi and coworkers\cite{Waka98} who found gap-less Goldstone modes, as expected.  In consequence, long-range order should be suppressed.  Therefore,  at finite temperature intra-edge spins correlations  are expected to survive up to a temperature dependent spin correlation length  that was computed by Yazyev and Katsnelson\cite{Yazyev08}.   At room temperature the magnetic correlation length was estimated to be 1$nm$. 

Within the mean field approximation the Fermi liquid picture of quasiparticles remains intact. However, in one dimension,
interactions are expected to fully renormalize
the one-particle bands,  and strongly correlated phenomena such as
spin-charge separation are expected to occur.   
This scenario has been considered for the case of quantum Hall
edge states\cite{Fertig06}, as well as the quantum Spin Hall 
edge states\cite{Sandler07} and more recently for the case of 
ferromagnetic zigzag edges as well\cite{Schmidt12}.

\section{Quantum Hall  phase}
In this section we discuss a totally different type of edge states.
 They arise in graphene when a magnetic field is applied perpendicular to its surface, they are extended much more than a few unit cells, and they can live in all types of edges, not only at zigzag boundaries. These edge states are associated to the 
so called Quantum Hall state, which \redmark{shows} a bulk
insulating behavior together with 
perfectly conducting  edge transport. In a classical naive picture,
bulk insulating behavior  is associated to electrons performing
closed orbits, whereas the edge electrons are able to move forward in the so called skipping orbits  by
bouncing on the interface. 

Nevertheless, the quantum mechanical treatment 
gives a much richer
understanding of QHE.   Bulk localization is associated to the emergence of 
the so called
\redmark{Landau levels (LL)}, with a discrete spectrum that replaces the energy bands and 
localized wave functions. The perfectly quantized  edge conductance can be understood by
taking into account the non-trivial topology
of the bulk states as well as the bulk to edge correspondence. 

The role of topology  can be seen as follows.  Both vacuum, or of that matter any trivial insulator,  and the quantum Hall state are insulators. However, there is no way to change the parameters in their Hamiltonians to turn one into the other without closing the band gap.  Therefore, the space of Hamiltonians generated by changing parameters in the trivial and quantum Hall insulators, without closing the gap, gives rise to   (at least) two disconnected classes of insulating Hamiltonians that can not be   deformed one into each other
without passing through a
\redmark{gapless} state. As a consequence, at the physical interface between two materials described with Hamiltonians that belong to these two different classes there must be a conducting state. 
The  general argument does not only apply
to the quantum Hall state, but also to a large amount of different
systems known as topological insulators. In the next sections
of this review, we shall discuss also some of them in the context of graphene.

\subsection{Electronic states}

\begin{figure}[hbt]
 \centering
  \includegraphics[width=0.5\textwidth]{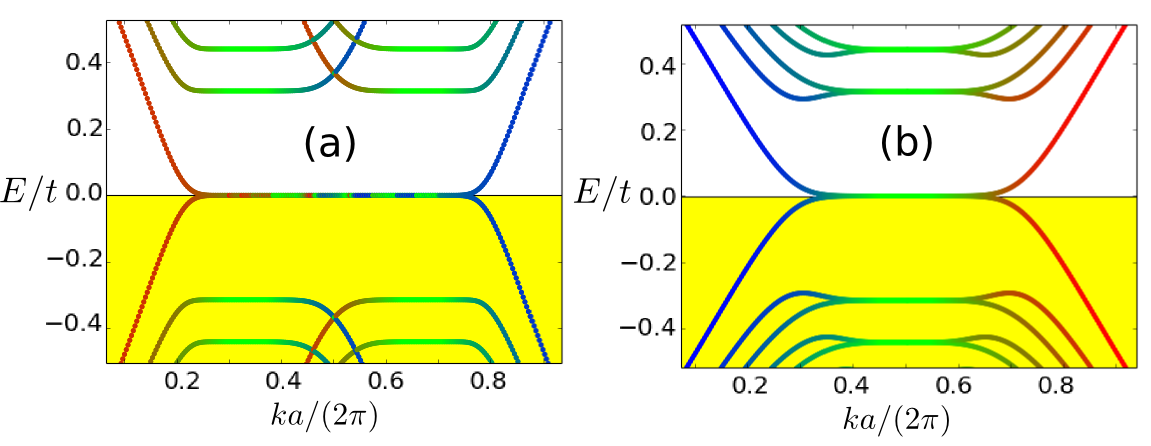}
\caption{LL for zigzag (a) and armchair (b) ribbons.
Band structure for a graphene ribbon with orbital magnetic field. The color code means red and blue for the edges and green denotes the bulk states.}
\label{f3}
\end{figure}

The description of the electronic states of graphene in a constant magnetic field starts with 
the choice of a vector potential $\vec{A}=B (-y,0,0) $ to be inserted in eq. (\ref{Peierls}), so that the original periodicity of the crystal is only preserved along the $x$ direction. As a result, the TB description of graphene under a magnetic field is very often restricted to one dimensional stripes  which permits to study the bulk and edge states on equal footing.  
\redmark{In a tight binding model
this minimal coupling is realized} via Peierls substitution. 
 The magnetic field introduces a new length scale in the problem: 
  \begin{equation}
  l_B=\sqrt{\frac{\hbar}{eB}}.
  \end{equation}
  which is $l_B= 25.7nm$ for $B=1T$.
    For a   sufficiently high  magnetic field the magnetic length is much smaller than the width of the system $W$. In that situation, 
  the bulk quantum states become localized, giving rise to LL, but  the edge states are dispersive, as shown in figure (\ref{f3})  obtained by numerical diagonalization of the TB Hamiltonian for a 1D stripe with zigzag edges (left) and armchair edges (right). 
  
  The spread of the edge wave functions depends on their energy with respect to their reference Landau level\cite{Lado13}, scales with  $l_B$ and is thereby much larger than the spread of the $B=0$ zigzag edge states discussed in the previous section.  Importantly, in the case of zigzag edges, both the magnetic edge states and the $E=0$ edge states coexist. Actually, inspection of the wave functions show that,  at a given valley,  the two bands with $E=0$ bands shown in fig.(\ref{f3}a)  correspond both to the $n=0$ Landau level in that  valley and the edge state. 
  
Further insight of these results can be obtained within  the so called   effective mass description, that can be formally derived from the $kp$ theory  of the energy bands. This technique  had tremendous importance in the description of semiconductors\cite{kpKohn}. In practical matters, it amounts to replace $\vec{q}$ in equation (\ref{kp}) by the momentum operator.
By so doing, we obtain an
 effective mass Hamiltonian   isomorphic to the Dirac Hamiltonian at each  valley:\cite{Semenoff84}
\begin{equation}
H_{\tau} = v_F \left( \Pi_x \sigma_x + \tau \Pi_y \sigma_y\right)  + \Delta\sigma_z,
\label{HAMIL-KP}
\end{equation}
where $\vec{\Pi}\equiv \vec{p}- e\vec{A}$ is the canonical momentum operator,  
$v_F=3 t a_{CC}/2 \hbar$,  
$\vec{\sigma}$ are the Pauli matrices describing the graphene sublattice degree of freedom and $\tau=\pm 1$ describes the valley index.

In the case of constant magnetic field perpendicular to the plane this Hamiltonian can be solved exactly.
Using the  transnational  invariance along the $x$ direction,  we can assume its eigenfunctions are products $e^{ik_x x}\vec{\phi}_{n}(k_x,y)$ which permits
replacing the operator $p_x$ by the quantum number $\hbar k_x$ in Eq. (\ref{HAMIL-KP}).   The Schrodinger equation becomes then a problem of two first order differential equations in one dimension:
 \begin{eqnarray}
 H_{\tau}=v_F (\hbar k_x-e B y)\sigma_x + \tau v_F p_y \sigma_y) + \frac{\Delta}{2}\sigma_z
 \label{Dirac-Landau-1}
\end{eqnarray} 

The solution of the problem is facilitated by introducing the dimensionless operators:
$Q(k_x)\equiv \left( \frac{y}{l_B}-k_x l_B \right)$ and $
P\equiv 
\frac{l_B}{\hbar} p_y,$
and the ladder operators
\begin{equation}
\alpha(k_x)=  \frac{1}{\sqrt{2}}\left(Q(k_x) + i  P \right),
\end{equation}
which  satisfy bosonic commutation relations
%
$[\alpha(k_x),\alpha(k_x)^{\dagger}]=1$. Using these operators, 
the Hamiltonian (\ref{Dirac-Landau-1})  for valley $\tau=-1$ can be written as: 
\begin{equation}
H= \frac{\Delta}{2}\sigma_z - \frac{\hbar\omega_0}{2\sqrt{2}}\left(\sigma^+ \alpha+ \sigma^-\alpha^{\dagger}\right)
\end{equation}
where $\sigma^{\pm}= \sigma_x \pm i\sigma_y$
where we have defined 
$\frac{ \hbar \omega_0}{2}
\equiv  \frac{\hbar v_F}{l_B}$.
 For $\tau=+1$ we have to exchange $\alpha$ and 
$\alpha^{\dagger}$.  This model is mathematically equivalent to the 
 Jaynes-Cummings model in quantum optics describing a two
level system coupled to a bosonic mode.  The resulting 
eigenfunctions and eigenvalues have the 
following properties  \cite{Koshino-Ando,Vasilopoulos2012,Lado13}.  First, 
there is a spectrum of states with wave functions 
with non-zero $\phi_A$ and $\phi_B$, with energies: 
\redmark{
\begin{equation}
E_{N}= \pm \sqrt{\Delta^2 + \frac{1}{2}\left(\hbar \omega_0\right)^2N}
\label{energies}
\end{equation}
}
with \redmark{$N$} a strictly positive integer. This spectrum 
comes with a twofold valley degeneracy, in addition to the 
twofold spin degeneracy.   Notice, that for $\Delta=0$ the 
Landau level spectrum for Dirac particles scales 
with $\sqrt{NB}$, in contrast to the linear scaling of 
Schrodinger particles, that can be retrieved 
in the limit of very large $\Delta$. 

In addition to the states described with eq. (\ref{energies}),  there 
is one {\em zero mode} per valley, with 
energy $E_0= \tau \frac{\Delta}{2}$ and with wave function 
fully sublattice polarized:
\redmark{
\begin{equation}
\phi_{\tau=+1}=
\left(\begin{array}{c} \phi_0 \\ 0 \end{array}\right), \,\,\,\,
\phi_{\tau=-1}
=\left(\begin{array}{c} 0 \\ \phi_0 \end{array}\right).
\end{equation}
}
For $\Delta=0$, these zero mode  would be 
degenerate at $E=0$. 
Importantly, the bulk Dirac-Landau 
levels  obtained analytically from the  effective mass approach 
are in very good agreement with the tight-binding 
calculation.    For the edge states it is also possible to work 
out the $kp$ theory\cite{Brey-Fertig06}, but this goes beyond 
the scope of this review.

\subsection{Topological origin of the edge states}



As we anticipated at the beginning of this section,
in the context of the
non trivial topology of the band structure,
the edge currents
are just protected interface states in a boundary between two insulators
with topologically inequivalent ground states.
Moreover, the perfect conductance of this state, is a result
of the relation between the quantum Hall conductance and the
so called Chern number
\begin{equation}
\sigma_{xy} = \frac{e^2}{\hbar}{\cal C}
\end{equation} 
where ${\cal C}$ is the flux over Brillouin zone 
\begin{equation}
{\cal C} =\frac{1}{2\pi} \int \Omega d^2k
\label{Chern}
\end{equation}
\redmark{
of the Berry curvature $\Omega$ 
\begin{equation}
\Omega =  i
\left (
\partial_{k_x} \langle \Psi | \partial_{k_y} | \Psi \rangle
-
\partial_{k_y} \langle \Psi | \partial_{k_x} | \Psi \rangle
\right )
\end{equation}
which acts as
an anomalous term in the velocity.
}

The Chern number of normal insulators  is ${\cal C}=0$.
In contrast,  the Chern number of the quantum Hall state is
an integer different from zero\cite{}, which results in the quantization of the Hall conductance.

In the case of interfaces between two different quantum Hall states,
the previous interpretation gives a simple way to predict the number
of edge states. For an interface between a first system
with Chern number ${\cal C}_1$
and a second system with Chern number ${\cal C}_2$, the number of
protected edge states
is simply $| {\cal C}_1-{\cal C}_2 |$.

\redmark{
In the quantum Hall regime,
the lack of edge states when a sublattice imbalance mass gap opens
at half-filling\cite{Abanin06,Fertig06,Lado14b},
accounts for 
${\cal C}=0$ and
can be rationalized in terms of adiabatic evolution
of the massive Dirac Hamiltonian from $B=0$.}
 Taking the half-filled case as a reference,  the   total Chern number away from half filling 
%
%
 can be calculated by adding  one unit for each Dirac Landau level  crossed by the Fermi energy as we move it up, including the valley and spin degeneracy.  Conversely, if we consider hole doping,  the Chern number becomes negative as we move the Fermi energy down
in energy.  \redmark{For example, starting from half filling
filling the electron $n=0$ Landau
level gives ${\cal C}=1$ per spin channel, as can be observed in Fig. 4.}
Analogously,  un-filling the
hole $n=0$ level gives a total ${\cal C}=-1$, per
spin channel. This will lead to remarkable
phenomena when
dealing with spin polarized systems. As we discuss below,
different fillings for the up and down channels,
can give rise to a very special type of \redmark{quantum}
spin Hall effect\cite{Abanin06}.

\section{Quantum Anomalous Hall phase } 
Another kind of topologically non-trivial quantized phase is 
the so called Quantum Anomalous Hall (QAH) phase.  This topologically insulating phase has properties analogous to those of the quantum Hall phase, including chiral edge states for which backscattering is impossible, and breaking of  time reversal symmetry.  However,  the QAH  phase is driven by internal degrees of freedom and interactions, such as spin-orbit coupling and exchange, rather than by the application of an external magnetic field.  
 The fabrication of materials showing this  phase has enormous potential since
they are expected
to show a perfect edge conductance as the quantum Hall state,
but without the drawback of having to apply large magnetic  fields\cite{}.

Two important graphene related  models present the QAH phase.  First,  the Haldane model for spinless fermions moving in a honeycomb lattice\cite{Haldane88}. Second,  graphene perturbed both with a  Rashba spin-orbit interaction 
 and subject to an  exchange field that couples to the off-plane spin component \cite{Qiao10,Qiao12}. 
  
  \begin{figure}[hbt]
 \centering
  \includegraphics[width=0.5\textwidth]{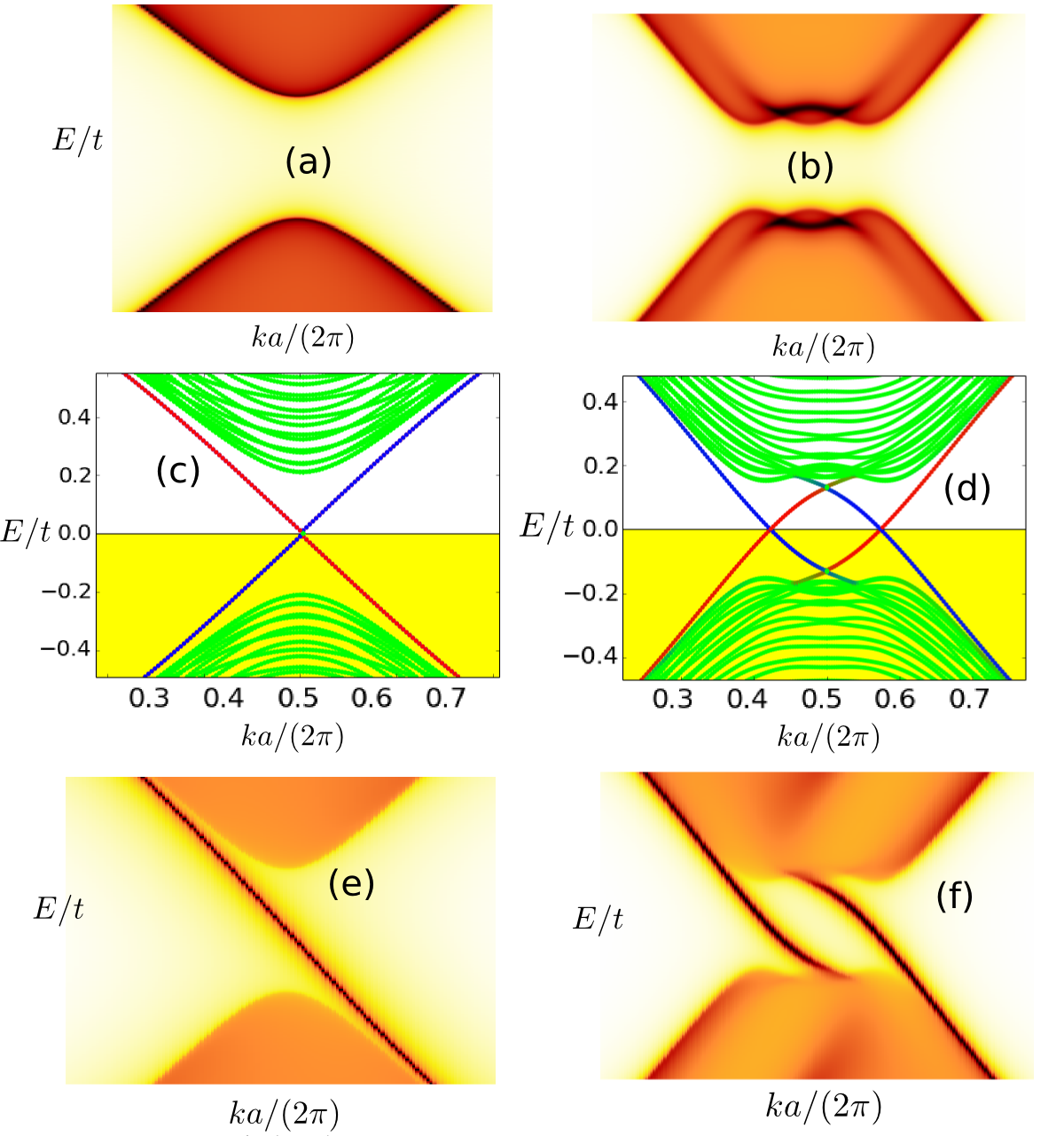}
\caption{Quantum Anomalous Hall phase in two toy models. Left panels: Haldane model\cite{Haldane88}. 
Right panels: model proposed by Qiao and coworkers\cite{Qiao10}.
Panels (a,b): Bulk density of states showing a gapped spectrum
resolved in the $k_x$ Panels (c,d):  band structure of an armchair ribbon system
for the same two models, showing that in addition to the gapped
bulk states, gap-less edge states show up. Panels (e,f):
density of states in the armchair edge of a semi-infinite plane,
showing a coexistence of bulk and one-way edge modes, calculated using the Green's function approach. 
}
\label{fig5}
\end{figure}
\subsection{Haldane model for QAH}
The toy model 
proposed by Haldane\cite{Haldane88} consists of
a honeycomb lattice in which there are local magnetic fields
 of different sign within the hexagon, but whose flux over it is zero.
This gives rise to an  imaginary second neighbor hopping, which
accounts for the local magnetic fields:
\begin{equation}
{\cal H}_{QAH1} = {\cal H}_0 +
t_2 \sum_{\langle \langle i,j \rangle \rangle}i \nu_{ij} c_{i}^\dagger c_{j}
\label{HH}
\end{equation}
where $\nu_{ij}=\left(\vec{d}_1\times\vec{d}_2 \right)\cdot\hat{z}$ is the chirality of the path of the second neighbor hopping, where 
   $\vec{d}_{1,2}$ are
defined as follows:   for a given pair of second neighbor atoms $i$ and $j$, with a common first neighbor $k$,     $\vec{d}_{1,2}$ are  the unit vectors along the bonds $ik$, $kj$, respectively. 

In this model, the bulk is characterized by a gapped spectrum as shown in
(Fig.\ref{fig5}a).  A $kp$ expansion of the $t_2$ term around the Dirac point would yield the following extra term in the Dirac Hamiltonian:
\begin{equation}
\delta  H = \frac{3}{2} \sqrt{3}t_2 \tau_z \sigma_z
\label{HH2}
\end{equation}
where the {\em mass } can be written as a valley dependent object, $\Delta_{QAH1}= 3\sqrt{3}\tau_z$, in contrast with eq. (\ref{kp}).  This has a very important consequence. 
 The  Chern number  associated to  Hamiltonian  (\ref{kp})
 reads\cite{RMP-Xiao}:
 \begin{equation}
{\cal  C}(\Delta,\tau_z)= \frac{1}{2}\tau_z \frac{\Delta}{|\Delta|}
 \end{equation}
  Therefore, in the Haldane model both valleys contribute with a Chern number with the same sign that sum ${\cal C}=1$, whereas in graphene with a trivial gap, opened with a term that breaks sublattice symmetry, the two valleys contribute with a Chern number with opposite sign, that give ${\cal C}=0$.

From the ${\cal C}=1$ for the Haldane model  we expect the existence of an
edge state.  This is seen in the spectrum of a ribbon with armchair edges as in-gap chiral state
 in (Fig.\ref{fig5}c), together with  the confined modes (in green).  The chiral character of the edge states prevents backscattering. Therefore, the Haldane model describes a phenomenology identical to the quantum Hall phase, but   without the presence of flat Landau
levels in the bulk.  In addition, a generalization of  the Haldane model played a crucial role in the proposal in the proposal of the quantum Spin Hall phase\cite{Kane-Mele1,Kane-Mele2}, as we discuss below. 

\subsection{Exchange plus Rashba model for QAH}
A more realistic model describing another realization of the 
 quantum anomalous Hall state in graphene was proposed by Qiao and coworkers\cite{Qiao10,Qiao12}.
In their model they describe electrons in graphene including both the
 Rashba spin-orbit term (eq. \eqref{Rashba})  and  exchange fields that splits the spin along the off-plane direction:  
\begin{equation}
{\cal H}_{QAH2} = {\cal H}_0+ {\cal H}_R+
\lambda\sum_{i,s} c_{is}^\dagger s_z c_{is} 
\label{QAH2}
\end{equation}
 A qualitative understanding of how these two terms open a gap in the bulk structure, can be gained from the following argument. The off-plane exchange field will split the Dirac cones\cite{Qiao10,Qiao12}, so that 
the $\uparrow$ and $\downarrow$ cones are degenerate on a circle in momentum space that, at half filling, corresponds to the Fermi circle.   The Rashba term couples $\uparrow$ and $\downarrow$, opening a band gap in the Fermi circle.

The $k_x$ resolved bulk DOS
shows again a gapped spectrum (Fig.\ref{fig5}b). However for a finite ribbon,
several bands crossing the Fermi energy
appear (Fig.\ref{fig5}d). From those four bands that appear in the
ribbon, two of them are located in one edge of the ribbon and two in the other.
This becomes clear  calculating the DOS \redmark{on} the edges, showing
also that the two bands present are
co-propagating, thus yielding a protection against back-scattering as in the
Haldane model. The Chern number of this is system
is ${\cal C}=2$,\cite{chern-graphene-qah} in agreement with the two states per edge
observed in Fig. \ref{fig5}d.

There are several proposals to modify graphene so that it is described by  Hamiltonian (eq. (\ref{QAH2})), including the use of 
ferromagnetic substrates,\cite{exp-qah-graphene} 
magnetic ad-atoms,\cite{qah-adatoms}
and even antiferromagnetic substrates\cite{graphene-qah-af}.
\redmark{
Finally,
a similar QAH can be obtained
in graphene
\cite{graphene-skyrmion}
by replacing the Rashba-like term by a topological
spin texture known as skyrmion.}


\section{Quantum Spin Hall phase}

The prediction\cite{Kane-Mele1,Kane-Mele2,Bernevig06}  and discovery\cite{Konig07} of the Quantum Spin Hall phase boosted the  research field  of topological insulators. Very much like the QH and QAH phases, the QSH phase has  an insulating bulk and conducting chiral edge states which, in contrast with the QH and QAH case,  have well defined spin chirality: a given edge hosts  two counterpropataging states with opposite spin projection, which has potential for spin electronics applications\cite{Pesin12,Han14}. Here we discuss two of the pioneer proposals for the QSH, both of them based on  graphene.  In the  first one, Kane and Mele proposed\cite{Kane-Mele1,Kane-Mele2}  that  intrinsic spin-orbit coupling would turn graphene into  QSH. Unlike the QH and QAH, this system would be time reversal invariant.  The second  proposal, by Abanin and coworkers\cite{Abanin06}  is rather special: ferromagnetic order of the  QH phase in graphene at half filling, for which ${\cal C}=0$, would results in a QSH phase, with ${\cal C}_{\uparrow}= -{\cal C}_{\downarrow}= +1$, but with time reversal symmetry clearly broken. We now review the most remarkable features of the edge states of these  two fascinating QSH states.


\subsection{QSH driven  by spin-orbit}
The influence of atomic  spin-orbit coupling on the  Dirac  bands, formed by $L_z=0$ atomic orbitals,  is subtle: to lowest order,  the 
atomic spin orbit coupling  $\lambda \vec{L}\cdot\vec{S}$ 
\redmark{within the p-manifold} has no effect at all. 
Second order interband scattering gives the first non-zero contribution but it
is not obvious how to include it in the Hamiltonian.
\redmark{Further work,\cite{graphene-dorb} showed that
inclusion of $d-channels$ lead
to a first order contribution in SOC.} In their
seminal work, based on symmetry considerations,  Kane and Mele postulated that the effective Hamiltonian for SOC in the subspace of the $\pi$ orbitals would be given by a spin dependent second neighbor hopping,  eq. (\ref{Hamil}), which turns out to be mathematically identical to the Haldane term that,   as discussed in the previous section,  turns spinless fermions into the  QAH  phase. Thus, the   Kane-Mele Hamiltonian ${\cal H}_0+ {\cal H}_{KM}$ for graphene with SOC is made of two decoupled
copies of the Haldane model, one per spin, with opposite sign of the gap-opening term.  Thus, the QSH phase described by the Kane-Mele model can be thought of as two QAH, one per spin, with opposite magnetization. 

In the neighborhood of the Dirac points, the KM model would give the following Hamiltonian: 
\begin{equation}
{\cal H}_{KM}(\vec{q})={\cal H}_{0}(\vec{q})
    +
 \frac{3}{2}\sqrt{3}t_{KM}  \tau s_{z}\sigma_{z} S_z
\label{Hamil-QSH}
\end{equation}
where ${\cal H}_{0}(\vec{q})$ is given by eq. (\ref{kp}). 
Thus,   the Kane-Mele gap-opening term respects both time-reversal symmetry
and inversion symmetry, in contrast with eq. (\ref{HH2}).

%
\begin{figure}
 \centering
   \includegraphics[width=0.5\textwidth]{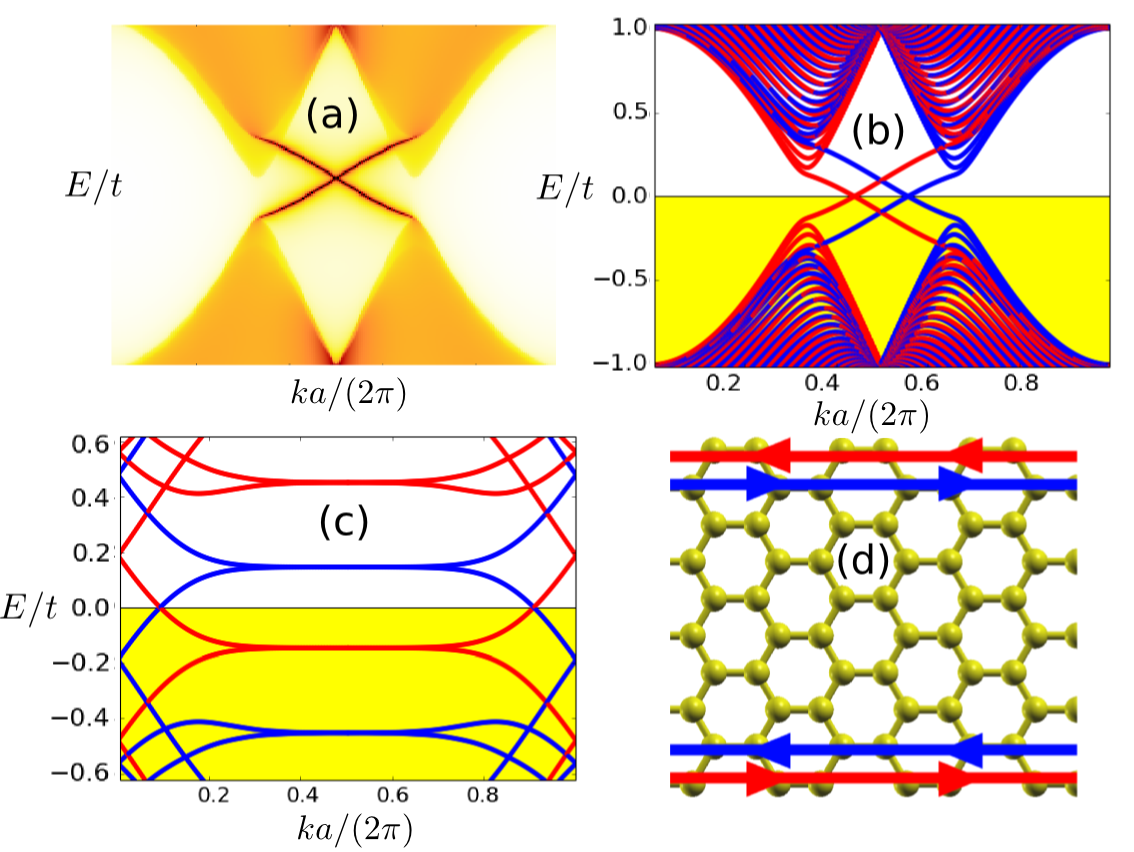}
\caption{
Top panels: Electronic band structure of a graphene zig-zag ribbon with a small staggered potential and an external magnetic field (off-plane) to resolve the degeneracies. Panel (a) shows the density of states at the edge of a semi-infinite plane with zig-zag termination. Panel (b) corresponds to the band structure of a zig-zag ribbon, the color represents spin polarization.  Panel (c):  the band structure for a ribbon in the presence of an off-plane Zeeman magnetic field and an orbital magnetic flux leading to the appearance of LL. In panel (d) a scheme of the QSH effect, where two spin polarized conducting states are present in each edge.}
\label{FIG6}
\end{figure}

 The fact that  the Kane-Mele model \redmark{consists}
of two copies of the Haldane model permits to anticipate its most salient electronic properties:  bulk states present a gap while the edges host one  chiral conducting states per spin channel. Since the sign of the gap is  given by the projection of the spin $S_z$, spin $\uparrow$ and $\downarrow$ edge states propagate in opposite directions, i.e., they have 
 a well defined  spin chirality. As a result, the  edge states carry an equilibrium spin current, in contrast with the QH and QAH phases for which edge states carry a net charge current (see figs. \ref{FIG6}(a,b,d)). 
 
Trivially, the Chern number of each spin channel is ${\cal C}_{S_z}=sgn(S_z)$, so that the spin Chern number ${\cal C}_{\uparrow}-{\cal C}_{\downarrow}=2$ would be finite but the the total Chern number would vanish. So, it would seem that the addition of spin-flip perturbations, such as the Rashba Hamiltonian, would result in intra-edge backscattering. However, this is not the case and it turns out that the Spin Hall phase is topologically different from the normal phase, even when $S_z$ is not conserved. To show this,   
Kane and Mele introduced
another topological invariant\cite{Kane-Mele2} in order to classify
these systems, the $Z_{2}$ invariant. The $Z_{2}$ group
consists only of two elements, say, $\{0,1\}$ and
according to Kane and Mele's classification, all
non-trivial insulators would have a $Z_{2}$ index
equal to $1$ while trivial insulator
would have $Z_{2}$ index equal to $0$.  When $Z_2=1$ the edge states are  counterpropataging Kramers pairs, so that only perturbations that break time reversal symmetry can produce backscattering. Importantly, Kane and Mele showed that the addition of a small Rashba term to their Hamiltonian would keep $Z_2=1$, and only when the Rashba term is large enough as to close the bulk gap would the system turn into a $Z_2=0$ trivial phase without topologically protected edge states. 

Whereas the observation of the spin-orbit coupling driven QSH phase in graphene is very difficult due to the small size of the spin-orbit gap (smaller than 0.1 meV\cite{Min06}), 
the observation of the Quantum Spin Hall  phase HgTe/CdTe
quantum wells \cite{Koing07} as well as in $InAs/GaSb$ inverted quantum wells\cite{Knez11}  has confirmed the existence of this fascinating class of materials.



\subsection{Quantum spin Hall effect without spin-orbit coupling}
Even if spin-orbit coupling can not drive graphene into the QSH phase, there
is \redmark{a} different mechanism that can do the trick and, contrary to other proposals\cite{Kane-Mele1,Kane-Mele2,Bernevig06}, it does not require spin-orbit coupling\cite{Abanin06}. 
This mechanism relies instead on the very special
Landau level structure that emerges upon application of a 
 magnetic field.

\redmark{Lets take} graphene  under the
influence of an off-plane magnetic field. At half filling,  the  Chern number of all the occupied bands is zero and two of the  four $n=0$ LL,  {\it valence} and {\it conduction} with spin degeneracy. 
\redmark{Due to} the four $n=0$  are degenerate, there are several alternative filling patterns. At half filling, all possible combinations give ${\cal C}=0$, but it is possible to have ${\cal C}_S=2$ if  we fill the two LL with the same spin, and  leave empty the other two.  Interestingly, this spin-polarized filling can be driven by Coulomb interactions\cite{Abanin06,Fertig06,Nomura06,Alicea06}  and also by intrinsic spin-orbit coupling\cite{Tabert13}. 


Thus, at half-filling, the spin-polarized quantum Hall phase has ${\cal C}=0$ and ${\cal C_S}=\pm 2$, so that it has the same topological 
properties than the QSH phase described by the Kane-Mele model.
\redmark{The energy bands corresponding to the addition
of a large Zeeman splitting} are shown in figure \ref{FIG6}(c). They feature a gapped bulk together with counter-propagating spin-filtered in-gap edge states.
Similar results can be obtained\cite{Lado14b} using a self-consistent calculation of the mean field Hubbard model, including an in-plane Zeeman field that favors ferromagnetic order.


Unlike the QSH phase driven by spin-orbit coupling,  spin-flip perturbations give rise to spin-flip edge backscattering and, because time reversal symmetry is broken,  it is not possible to accommodate this QSH system into the $Z_2$ classification of Kane and Mele.  Interestingly, there is very strong experimental evidence that this interaction driven QSH phase has been observed in the experiments, where the ferromagnetic order is favored by applying a quite large in-plane magnetic field\cite{Young14}.  At half filling, the experiments show how  two terminal conductance of graphene is tuned from 0, corresponding to a gapped quantum Hall phase  to 1.8$\frac{e^2}{h}$,  that most likely is the QSH phase predicted\cite{Abanin06} by Abanin and coworkers.

\section{Coulomb driven breakdown of the edge conduction in the QSH phase}
The gapless edge states in the quantum spin Hall states
are protected by the spin Chern number in the case
of magnetic field driven phase,\cite{Abanin06} and by the $Z_2$ invariant
in the case of the SOC driven.\cite{Kane-Mele2} These topological invariants
are well defined as long as $S_z$ is a good quantum number (Spin Chern)
and time reversal symmetry is present ($Z_2$).
In the case spontaneous magnetism shows up,
such conditions no longer apply, so the gap-less edge states
are no longer guaranteed.

In this section we show two examples of how Coulomb interactions can lead to the 
disappearance of spin chiral edge states
due to the emergence of spontaneous edge magnetization. Following the structure of the previous section,  we consider both the SOC driven, and the  quantum Hall ferromagnet QSH phases.

\subsection{Kane-Mele-Hubbard model}
In order to study the interplay between edge magnetism and topologically protected edge states we use the so called Kane-Mele-Hubbard model\cite{Rachel10,Soriano10,kmu1,kmu2,kmu3,kmu4,Lado14a}
Although the gapless edge states are protected against
time reversal perturbations, an important issue
is whether electron-electron interactions
are able to drive the system into a symmetry breaking state.
The study of the bulk properties of the Kane-Mele-Hubbard model has attracted considerable attention\cite{kmu1,kmu2,kmu3,kmu4}. Here we focus on the competition between SOC and magnetic order at the edges\cite{Soriano10,Gosalbez12,Lado14a}.

\begin{figure}
 \centering
                \includegraphics[width=0.5\textwidth]{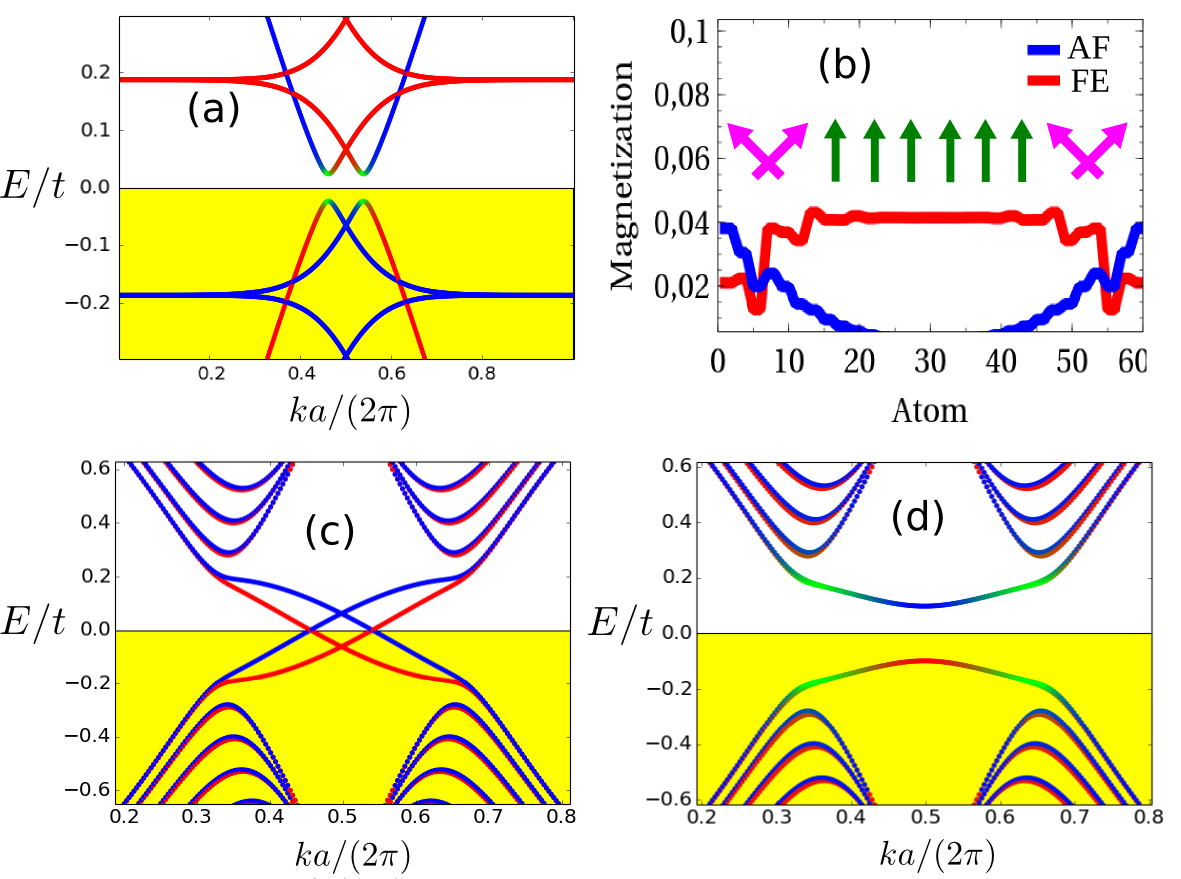}
\caption{Band structure (a) and local magnetization (b) of the ferromagnetic
quantum Hall state with interaction driven edge canted antiferromagnetism.
The local edge antiferromagnetism, gapes out the original counter
propagating spin filtered edge state.
Band structure of a zigzag ribbon with spin orbit, whose edge magnetic moments
driven by interaction are of plane (c) and in-plane (d). The in-plane
magnetization destroys the gapless states due to breaking of
TR and spin mixing, whereas the off-plane does not because $s_z$ is
still conserved.
}
\label{fig7}
\end{figure}

Edge magnetic order would break  time reversal
symmetry and could gap out those edge states. 
Early work assumed that magnetic order would be off-plane\cite{Soriano10,Gosalbez12}.  In this situation, 
the gapless edge states survive  (Fig.\ref{fig7}c), at least for small $U$ and large SOC\cite{Soriano10}. 
In general,   magnetic order in a topological insulator
protected by time reversal symmetry should destroy the gapless edge states.  This case is exceptional because
 in addition
to being protected by time reversal symmetry, the Kane-Mele model
shows a well defined spin Chern number, which relies
only in conservation of $S_z$, preserved by the off-plane magnetic order. 

The situation changes radically  when the edge magnetic moments
lie in-plane.\cite{Rachel14,Lado14a} Then, $S_z$ is no longer a good quantum number,
and thus the protection that remained in the off-plane magnetic case
no longer holds.
Now, the in-plane magnetism
opens up a spin mixing channel, 
driving the system into
an edge insulator, as shown in Fig.\ref{fig7}d. 

Interestingly,  spin orbit coupling produces magnetic anisotropy, so that 
both configurations are not expected to be energetically equivalent. 
Actually,  the ground state of the mean field Hubbard model  is in fact
the in-plane magnetic state.\cite{Lado14a}
However, the disappearance of the chiral edge states can
be restored, by overcoming the magnetic anisotropy of the system,
and forcing the states to lie off-plane. This could be achieved 
by a magnetic substrate, which if made switchable, would allow to
control the flow of the spin channels along the edge at will.


\subsection{Magnetic field driven QSH phase}
In this situation, the shifting between the LL driven by the
magnetic field yields the effective Quantum Spin Hall state.

The state arises when the 
in-plane field is large enough to overcome
the possible antiferromagnetic state.\cite{herbutqh}
However two situations of edge states are possible. On one hand,
at weak interactions the edge becomes ferromagnetic at the same time
as the bulk. On the other hand, when interactions are stronger,
the edge of the system is able to develop a canted magnetic order,
even though the bulk remains ferromagnetic.

Experimentally, it was observed that in order to reach
the QSH in graphene, a large magnetic field
of 20 T had to be applied.\cite{Young14}
Two different scenarios are possible to understand
the large field needed to overcome a the fully gapped phase. 
The first one
consists on having an AF ground state at zero in-plane field,
turning it into ferromagnetic above the
critical field.\cite{herbutqh} In this
situation, the edge does not play a key role, and
the conductance only depends on the bulk magnetic order.
A second scenario appears when considering that, according to
theoretical predictions,\cite{Nomura06,Alicea06} the order in the bulk is
actually
ferromagnetic due to the long range Coulomb interaction.
Starting in this situation, two counter-propagating spin
polarized states
appear on the edge of the system.\cite{Abanin06} 
Those states, can be highly vulnerable to the local electron-electron
interactions, and
a local canted antiferromagnetic gap can be opened, gaping out the states
and destroying the quantum Hall state. 
Thus, the critical field needed
to reach the QSH is the one needed to recover ferromagnetism on the
edge.   
In the context of the
Hubbard model, it is found that at medium interactions $(U=2t)$,
even in the presence of an exchange field, the bulk of the system can be
in a ferromagnetic state, at the same time the edge remains a
canted antiferromagnet, as shown in Fig.\ref{fig7}b.


%
%
%

 \section{Experimental evidence}
 All the previous discussion was
focused on our theoretical understanding of a variety
of different edge states in graphene and graphene like systems.  Here we briefly touch upon the experimental evidence that has been obtained so far.

 \subsection{Zigzag Edge states and edge magnetism}
   We first discuss the observation of zigzag  edge states  and edge magnetism.   Scanning tunneling microscopy (STM) had permitted early on to
 identity zigzag and armchair edges in graphite\cite{Kobayashi05}.
  A fairly convincing evidence\cite{Crommie2011} of the existence of localized edge states at chiral edges of ultrahigh quality graphene ribbons obtained by unzipping carbon nanotubes\cite{Dai08} was obtained by means of scanning tunneling microscopy.   The same group also showed the capability to engineer the edges by means of plasma etching\cite{Zhang13} of graphene nanoribbons (GNR). They found that
 hydrogen-plasma-etched GNRs are generally flat, free of structural reconstructions, and terminated by hydrogen atoms with no rehybridization of the outermost carbon edge atoms. Both zigzag and chiral edges show the presence of edge states.  Edge states at the interface between graphene and h-BN  have also been imaged with STM\cite{Drost14}.   
   
The direct evidence for edge magnetism is less abundant, and most of the claims are based on  indirect  evidence.  Part of the problem is that conventional magnetic probes, such as SQUID and EPR,  have a resolution limit in the range of 10$^{12}\,\mu_B$ that would require graphene flakes beyond current capabilities\cite{Nair12}.  Therefore, such measurements are performed on solutions of graphene processed by various techniques, including  unzipping carbon nanotubes whose grow very often requires ferromagnetic catalyst. Therefore, the observation of room temperature ferromagnetism in graphene ribbons\cite{Rao12} should be revised under the light of the recently work \cite{Sepioni12b} discussing artifacts for  analogous claims for highly oriented pyrolitic  graphite\cite{graphite-artifacts}. 

These shortcomings highlight the   need to study edge magnetism in graphene using  some sort of local probe, preferably spin polarized STM\cite{Wiesendanger09}, or some type of nanomagnetometry technique, such as NV center detectors, that can be used to probe local moments on a surface\cite{Ternes14}.  A non-magnetic local probe was used by Tau and coworkers\cite{Crommie2011} who compared their fairly extensive STM spectroscopy, both along the edge of the ribbon and in the direction perpendicular to the edge with mean field Hubbard model calculations  where edge magnetism was included.  The good agreement between theory and experiment is certainly a hint that edge magnetism was present. Further computational work using density functional calculations have confirmed\cite{Mazzarello13} the presence of edge magnetism in hydrogenated graphene ribbons deposited on Au(111).
However, they also showed edge magnetism would not survive on 
other substrates, such as Cu(111) and Ag(111). 

The recent claims of room temperature edge ferromagnetism in graphene ribbons\cite{Magda14} are also based on STM spectroscopy and comparison of the data to well established theoretical results.  Unlike the work of Tau et al\cite{Crommie2011},  a detailed evolution of the  $I(V)$ curves along the ribbons is not provided.  In addition, edge magnetism is definitely not expected to survive up to  room temperature\cite{Yazyev08}, although the effect of long-range order depletion due to quantum and thermal fluctuations on the spectral function of the quasiparticles has not been studied.  A local probe of magnetism would be desirable in order to back up these extraordinary claims.

The recent report\cite{Baringhaus14}  of ballistic  transport with a conductance of $G=\frac{e^2}{h}$ in epitaxial graphene nanoribbons would imply time reversal symmetry breaking compatible with edge magnetism.  Great care was taken by the authors of this paper to rule out a great number of possible experimental artifacts and the understanding of their data remains a very interesting challenge for theory. 

The finite spin of  small triangulate molecules, with $S=\frac{1}{2}$ and $S=1$ has been observed experimentally, in macroscopic samples, using electron paramagnetic resonance techniques.  

\subsection{Edge transport in topological phases in graphene like systems}

 The observation of the quantum Hall effect in graphene with perfect quantization of the Hall conductance provides an extremely convincing, albeit indirect, evidence of the existence of Quantum Hall edge states. The  Quantum Spin Hall  predicted by Abanin {\em et al.}\cite{Abanin06} has been recently found experimentally\cite{Young14} by A. Young and coworkers.  They verified, by means of local capacitance measurements, the opening of a band-gap in bulk and the edge nature of the transport by means of a floating gate technique\cite{Young14}. The edge conductance in the QSH-like phase was $G=1.8 G_0$, below the ideal case of $G=2G_0$ which suggest that some sort of spin-flip backscattering mechanism is present in the sample. 
 
 \section{Conclusions and outlook}
 We have provided an overview of the electronic properties of edge states in graphene, and related materials, using as a guideline the standard tight-binding model for electrons in a honeycomb lattice.  Remarkably, small additions on  the same model  that describes the main properties of graphene, including the observed quantum Hall phase,  have led to the prediction of  the Quantum Spin Hall and Quantum Anomalous Hall phases,  inspiring the search of graphene based topological phases.  Therefore, small modifications of graphene, via proximity or functionalization, are expected to drive graphene into these exciting topological phases.  
  
Edge states play an important role in the study of some of the most fascinating  electronic properties of graphene and graphene like materials. 
 The study of graphene edge states has ramifications in many branches of modern material science, including graphene spintronics\cite{},
 organic chemistry\cite{},  quantum topological phases\cite{}, and metrology\cite{},  just to mention a few. 

Combined with the discovery of new graphene-like materials, and the construction of artificial structures where different two dimensional crystals are stacked and put in close proximity to materials with various types of electronic order, such as ferromagnetism  and superconductivity,   we can expect that the study of the electronic properties of edge states will bring many exciting results in the years to come.


 JFR acknowledges  financial supported by MEC-Spain (FIS2010-21883-C02-01) 
  and Generalitat Valenciana (ACOMP/2010/070), Prometeo. This work has been financially supported in part by FEDER funds.  We acknowledge financial support by Marie-Curie-ITN 607904-SPINOGRAPH. J. L. Lado and N. Garc\'ia thank the hospitality of the Departamento de F'sica Aplicada at the Universidad de Alicante.

\end{document}